\documentclass[11pt]{article}
\usepackage{axodraw,mathrsfs}

\catcode`\@=11
\@addtoreset{equation}{section}
\makeatother

\title{
\centerline{\normalsize hep-ph/0105305 \hfill SINP/TNP/01-12}\bigskip
\bf Rephasing-invariant $CP$ violating parameters with Majorana
neutrinos} 

\author{\bf Jos\'e F. Nieves\footnote{nieves@ltp.upr.clu.edu}\\ 
Laboratory of Theoretical Physics, Department of Physics\\ 
P.O. Box 23343, University of Puerto Rico\\ 
R\'{\i}o Piedras, Puerto Rico 00931-3343\\ 
\and 
\bf Palash B. Pal\footnote{pbpal@theory.saha.ernet.in}\\ 
Abdus Salam International Centre for Theoretical Physics\\ 
Strada Costiera 11, 34100 Trieste, Italy\\ 
and\\
Saha Institute of Nuclear Physics\\ 
1/AF Bidhan-Nagar, Calcutta 700064, India\thanks{Permanent address}}

\date{May 2001}

\begin{document}
\maketitle

\begin{abstract}
We analyze the dependence of the squared amplitudes
on the rephasing-invariant $CP$-violating parameters of the lepton
sector, involving Majorana neutrinos, for various  
lepton-conserving and lepton-violating processes.
We analyze the conditions under
which the $CP$-violating effects in such processes vanish, in terms of
the minimal set of rephasing invariants, giving special attention
to the dependence on the extra $CP$-violating parameters that
are due to the Majorana nature of the neutrinos. 
\end{abstract}

\section{Introduction}
Sometime ago \cite{Nieves:1987pp}, we presented a rephasing-invariant
formulation of $CP$ violation in the standard model, paying particular
attention to the $CP$ violation in the lepton sector involving
Majorana neutrinos. In that reference we generalized the previous work
by Greenberg \cite{Greenberg:1985mr}, and by Dunietz, Greenberg and Wu
\cite{Dunietz:1985uy}, on the quark sector.  We offered a well-defined
prescription for identifying a minimal set of rephasing-invariant
$CP$-violating parameters for any number of generations of Dirac
fermions, which is applicable to the quark sector, and the method was
extended to the lepton sector with Majorana neutrinos.

However, the question of how those rephasing invariants enter in the
amplitudes or rates for physical processes was touched upon only
briefly there. It is certainly of interest to see how different
physical quantities may depend on the rephasing invariants in a given
physical process, since this is necessary in order to understand the
conditions under which the $CP$-violating effects in a given process
may vanish. 

The question is more pressing when we consider the case of Majorana
neutrinos.  In this case, there are more independent rephasing
invariant $CP$-violating parameters than in the quark sector or in a
lepton-number conserving lepton sector with the same number of
families, and are due to the Majorana nature of the neutrinos
\cite{Bilenky:1980,Schechter:1980gr,Doi:1981yb}. Accordingly, there is
a larger number of rephasing invariants for this case.
It is generally believed that these parameters
show up only in lepton number violating processes
\cite{Schechter:1981gk}, and the rephasing invariant formulation has
proven to be useful in analyzing a variety of such processes
\cite{O'Donnell:1995ak,Rangarajan:1998hj,Liu:2001xs,Bilenky:2001rz}.  A
particularly important question is whether those additional parameters
can show up in lepton number conserving processes as well.

Another interesting question arises from the following.  In the
neutrinoless double beta decay process, the $CP$-violating terms in
the amplitude squared consist of products of the real and the
imaginary part of the relevant rephasing invariant parameters
\cite{O'Donnell:1995ak,Bilenky:2001rz,Aguilar-Saavedra:2000vr}.
Thus it would seem, at least in that context, that the potential
observability of any effect associated with the $CP$-violating part of
the rephasing invariants requires also that the real part of those
parameters be non-zero.  Therefore, an important question to ask is
whether this is a general feature, of this and other processes, or
whether it is possible to have observable effects due to the
$CP$-violating part of the rephasing invariants even if their real
parts are zero.

To this end, we consider various generic physical processes,
classified according to whether they conserve total lepton number, or
by how many units they violate it.  We find their generic dependence
on the rephasing-invariant $CP$-violating parameters, from which the
conditions under which the $CP$-violating effects may vanish follow.
This procedure allows us to consider the questions that we have posed,
and opens the way to make more general statements, divorced from the
kinematical aspects of the processes considered.

In Section\ \ref{sec:formalism} we introduce the notation, conventions
and other ingredients needed for our analysis.  They are used in
Section\ \ref{sec:analysis} to consider various classes of processes
with regard to the questions mentioned above and, finally, our
conclusions are summarized in Section\ \ref{sec:conclusions}.
 
%
%
\section{Basic vertices and amplitudes}
\label{sec:formalism}
\subsection{The rephasing invariants}
\label{sec:rephinv}
As shown in Ref.\ \cite{Nieves:1987pp}, a suitable set of
rephasing invariants to describe $CP$ violation in the lepton sector
with Majorana neutrinos, is constructed out of the products
\begin{equation}
\label{sinv}
s_{\alpha i j} = V_{\alpha i}V^\ast_{\alpha j} K^\ast_i K_j \,,
\end{equation}
where the $V_{\alpha i}$ are the elements of the mixing matrix 
that appears in the charged-current part of the Lagrangian,
\begin{eqnarray}
{\mathscr L_{\rm cc}} = \frac{g}{\sqrt{2}} W_\lambda 
\sum_{\alpha,i}
V_{\alpha i} \overline\ell_\alpha \gamma^\lambda L \nu_i 
+ \mbox{H.c.} \,,
\label{Lcc}
\end{eqnarray}
while the $K_i$ are the phases that appear in the Majorana
condition\footnote{We take this opportunity to note that Eq. (16) of
Ref.~\cite{Nieves:1987pp} contains a typographical error.  The correct
formula, consistent with the notation of that paper, is given here in
Eq.\ (\ref{majcond}). We thank Serguey Petcov for bringing this to our
atention.}
\begin{eqnarray}
\label{majcond}
\nu^c_i = K_i^2 \nu_i \,.
\end{eqnarray}
We have denoted by $\ell_\alpha$ and $\nu_i$ the charged lepton and
the neutrino mass eigenfields, respectively, and $L =
\frac12(1-\gamma_5)$. The sum in Eq.\ (\ref{Lcc}) goes over all
charged leptons and all neutrino eigenstates.

For processes which do not involve lepton number violation on any
fermion line, it is convenient to use the rephasing invariant
combinations\footnote{In connection with these and later formulas in
this paper, it should be remarked that the flavor indices, both for
charged leptons and neutrinos, are \underline{not} assumed to be
summed over whenever they are repeated. Summation is always explicitly
indicated wherever applicable.}~\cite{Greenberg:1985mr,Dunietz:1985uy}
\begin{eqnarray}
t_{\alpha i\beta j} = V_{\alpha i} V_{\beta j} V^*_{\alpha j}
V^*_{\beta i} \,.
\label{tinv}
\end{eqnarray}
In these cases, the independent $CP$-violating parameters for $N$
lepton generations can be taken to be
\begin{equation}
\label{minimalt}
\mbox{Im}(t_{\alpha i eN})\,, \quad \alpha\le i\,,\quad 
\alpha \not = 1\,,\quad i \not = N\,.
\end{equation}

For processes that violate lepton number, there are two ways 
to proceed\cite{Nieves:1987pp}.  One way is to take the minimal set
given in Eq.\ (\ref{minimalt}) and append it with the set
\begin{equation}
\label{minimalsadd}
\mbox{Im}(s_{1iN})\,, \quad i \not = N\,.
\end{equation}
An alternative is to use the fact that the $t$-invariants can be
expressed in terms of the $s$-invariants as
\begin{eqnarray}
\label{tsrelation}
t_{\alpha i\beta j} = s_{\alpha ij} s^*_{\beta ij} \,,
\end{eqnarray}
so that we only need to make reference to the $s$-invariants.  If we
follow this option, a minimal set of independent $CP$-violating
parameters, which are rephasing invariant, are
\begin{equation}
\label{minimals}
\mbox{Im}(s_{\alpha i N}) \,, \qquad \alpha \leq i \,, \quad i \not = N \,.
\end{equation}
For example, with this choice, the three independent rephasing
invariants in the case of three families of leptons are
\begin{equation}
s_{113}\,,\; s_{123}\,,\; s_{223} \,.
\end{equation}
while in the first approach, the set is
\begin{equation}
\label{choice13gen}
s_{113}\,,\; s_{123}\,,\; t_{2213} \,.
\end{equation}
%

\subsection{Basic amplitudes}
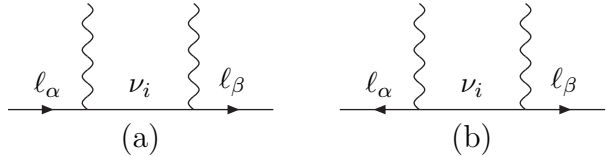
\begin{figure}[hbtp]
\begin{center}
\begin{picture}(100,60)(0,-15)
\ArrowLine(0,0)(30,0)
\Text(15,5)[b]{$\ell_\alpha$}
\Line(30,0)(70,0)
\Text(50,5)[b]{$\nu_i$}
\ArrowLine(70,0)(100,0)
\Text(85,5)[b]{$\ell_\beta$}
\Photon(30,0)(30,40)24
\Photon(70,0)(70,40)24
\Text(50,-5)[t]{\large (a)}
\end{picture}
\qquad
\begin{picture}(100,60)(0,-15)
\ArrowLine(30,0)(0,0)
\Text(15,5)[b]{$\ell_\alpha$}
\Line(30,0)(70,0)
\Text(50,5)[b]{$\nu_i$}
\ArrowLine(70,0)(100,0)
\Text(85,5)[b]{$\ell_\beta$}
\Photon(30,0)(30,40)24
\Photon(70,0)(70,40)24
\Text(50,-5)[t]{\large (b)}
\end{picture}
\end{center}
\caption{The two basic subdiagrams for processes that do not
involve neutrinos in the external states. Diagram (a) conserves
total lepton number, while Diagram (b)
changes it by 2 units.}\label{f:blocks}
\end{figure}

We consider processes such as $nn\to ppee$ (neutrinoless double-beta
decay) and $\mu\to 3e$, in which the individual
lepton flavors are not conserved, and which do not involve neutrinos
in the external states. Since the Standard Model conserves baryon
number, and since the total number of external fermions in any given
process must be even, then the number of external charged leptons must
be even also.  This means that, in such processes, the total lepton
number is either conserved, or it is violated in multiples of two
units. Since the only source of flavor violation in the Standard Model
is the charged current, it follows that any process of the kind we are
considering must contain the basic subprocesses shown in
Fig.~\ref{f:blocks} as building blocks.  Apart from the kinematical
factors, the amplitudes associated with these two subprocesses involve
the following elements of the neutrino mixing matrix
\begin{eqnarray}
\label{A}
A_{\beta\alpha i} & = & V_{\beta i} V^*_{\alpha i} \\
\label{B}
B_{\beta\alpha i} & = & V_{\beta i} V_{\alpha i}(K^\ast_i)^2  \,,
\end{eqnarray}
corresponding to the diagrams \ref{f:blocks}a and \ref{f:blocks}b,
respectively.

The appearance of the $(K^\ast_i)^2$ factor in Eq.\ (\ref{B})
originates in the contraction of the form
$\left<0\left|\nu_i\nu_i\right|0\right>$ that appears in the Wick
contraction of the interaction terms, which can be turned into the
neutrino propagator $\left<0\left|\nu_i\overline\nu_i\right|0\right>$
with the help of the Majorana condition of Eq.\ (\ref{majcond}).  An
alternative treatment, which amounts to a redefinition of the neutrino
field, is to associate a factor of $V_{\alpha i}K^\ast_i$ with each
$W$ vertex in a diagram.  In any case, the underlying reason is that,
since the processes do not involve neutrinos in the external states,
then the amplitudes (and not just the rates) must be invariant under a
rephasing of the neutrino field alone, which requires that the mixing
matrix elements appear in the combination $V_{\alpha i}K^\ast_i$.

The precise way in which the factors $A_{\beta\alpha i}$
and $B_{\beta\alpha i}$, and/or their products, may
enter in the total amplitude for a process depends on the particular
process under consideration. We consider such processes in the
following section,
and identify the $CP$-violating parts of their amplitudes
in terms of the rephasing invariants.  

%
%
\section{Analysis of specific processes}
\label{sec:analysis}
In this section, we study how the amplitudes for various
processes depend on the parameters $s_{\alpha i j}$.  As we will see,
there are processes for which the $CP$ violating terms vanish if
either the real or the imaginary part of the relevant $s$-invariants
are zero. However, this result is not general, and in particular it
does not hold if we consider higher order corrections. In the general
case, the $CP$ violating terms do not vanish when the real parts of
the $s$-invariants are zero, although, as expected, they do vanish
when the imaginary parts of the $s$-invariants are zero.  Furthermore,
there are processes that conserve lepton number for which the $CP$
violating terms depend on the $s$-invariants and cannot be expressed
in terms of the $t$-invariants alone. 

\subsection{Classification scheme}
\label{sec:classification}
In order to organize the analysis, we divide the processes into two groups,
according to whether the total lepton number changes or not.
These two groups are further subdivided into smaller groups
depending on the change of the individual lepton
flavors, $\Delta L_{\alpha}$. It is then useful to introduce the
following device.  In a given diagram, the subamplitude $A_{\beta\alpha i}$ 
increases the lepton flavor $\beta$ and decreases the flavor $\alpha$,
both by one unit, and leaves unchanged all the other flavors.
To use this feature in a precise way, we introduce a flavor counting operator
$L_\gamma$, for each flavor $\gamma$, and write
\begin{equation}
\label{LA}
L_\gamma\{A_{\beta\alpha i}\} = \delta_{\gamma\beta} - 
\delta_{\gamma\alpha} \,.
\end{equation}
Similarly, 
\begin{equation}
\label{LB}
L_\gamma\{B_{\beta\alpha i}\} = \delta_{\gamma\beta} + 
\delta_{\gamma\alpha} \,,
\end{equation}
and the total lepton number operator is defined as
\begin{equation}
L = \sum_\gamma L_\gamma
\end{equation}
where the sum runs over all the lepton flavors. This definition,
together with Eqs.\ (\ref{LA}) and (\ref{LB}) imply
\begin{eqnarray}
L\{A_{\beta\alpha i}\} & = & 0 \nonumber\\
L\{B_{\beta\alpha i}\} & = & 2 \,.
\end{eqnarray}
These definitions are complemented by the rules for 
the complex conjugate amplitudes
\begin{eqnarray}
L_\gamma\{A^\ast_{\beta\alpha i}\} & = & - 
L_\gamma\{A_{\beta\alpha i}\} \nonumber\\
L_\gamma\{B^\ast_{\beta\alpha i}\} & = & - 
L_\gamma\{B_{\beta\alpha i}\} \,,
\end{eqnarray}
which imply
\begin{eqnarray}
L\{A^\ast_{\beta\alpha i}\} & = & 0 \nonumber\\
L\{B^\ast_{\beta\alpha i}\} & = & -2 \,.
\end{eqnarray}

The amplitude $M$ for any given process is a sum of terms
\begin{equation}
M = \sum_{X = 1}^N M_X f_X \,,
\end{equation}
where the $f_X$ are kinematical factors and
each $M_X$ is made of a product of the subamplitudes
$A_{\beta\alpha i}$, $B_{\beta\alpha i}$ and/or their complex
conjugates. We indicate this schematically by writing 
\begin{equation}
M_X = C_1 C_2 ... C_{N_X} \,,
\end{equation}
where each $C_x$ stands for one of the subamplitudes or its complex
conjugate.  The flavor counting rules are amended to 
include the rules for the product
\begin{equation}
L_\gamma\{M_X\} = \sum_{x = 1}^{N_X} L_\gamma\{C_x\} \,.
\end{equation}
When building the amplitude for a given process out of products of the
subamplitudes $A_{\beta\alpha i}$, $B_{\beta\alpha i}$ (and their
complex conjugates),
the basic requirement is that all the terms $M_X$ for that process
must have the same lepton flavor; i.e.,
\begin{equation}
L_\gamma\{M_1\}= L_\gamma\{M_2\} = ... = L_\gamma\{M_{N}\}
\end{equation}
for all flavors $\gamma$. 

\subsection{$\Delta L = 2$ processes}
We subdivide this group of processes according to whether only
one flavor changes or two of them change.

\subsubsection{One flavor change}
In this kind of process, the amplitude is such that,
for some flavor $\alpha$,
\begin{eqnarray}
\label{case1}
L_\alpha\{M\} & = & 2 \nonumber\\
L_\gamma\{M\} & = & 0 \qquad (\gamma \not = \alpha) \,.
\end{eqnarray}
An example of this kind of process is neutrinoless double beta decay, or
simply $W^-W^-\to e^-e^-$,
which is of the type given in Fig.~\ref{f:blocks}b, with 
$\ell_\alpha = \ell_\beta = e$.  

It is then easy to see that, in the notation of
(\ref{B}), the amplitude for this type of process is of the form
\begin{eqnarray}
{\cal M}  = \sum_iB_{\alpha\alpha i} f_i \,,
\end{eqnarray}
where $f_i$ are functions that depend on the kinematic variables
of the problem, including the mass of the internal neutrino
line. Thus,~\cite{O'Donnell:1995ak}
\begin{equation}
\Big|{\cal M}\Big|^2 =
\sum_{ij} \left( B_{\alpha\alpha i}f_i\right) 
\left( B_{\alpha\alpha j}f_j\right)^\ast 
= \sum_{ij} \Big( s_{\alpha ij} \Big)^2 f_i f_j^\ast \,,
\end{equation}
where we have used Eqs.\ (\ref{sinv}) and (\ref{B}).
Therefore, any $CP$ violating effect is proportional to the imaginary part, 
$\mbox{Im}\;\delta_{\alpha\alpha ij}$, of the interference
terms
\begin{eqnarray}
\delta_{\alpha\alpha ij} 
%
& = & \left(s_{\alpha ij}\right)^2\,.
\label{eeij}
\end{eqnarray}
Since $\mbox{Im}\, \left(s_{\alpha ij}\right)^2 = 2(\mbox{Re}\,
s_{\alpha ij})(\mbox{Im}\, s_{\alpha ij})$,
then the following statement can be made for this type of process:
any $CP$ violating effect 
due to a non-zero imaginary part of an element
$s_{\alpha ij}$ vanishes unless
the real part of the same element is non-zero.

\subsubsection{Two flavors change}
\label{subsec:case2}
In this kind of process the amplitude is such that, for some two
flavors $\alpha,\beta$,
\begin{eqnarray}
\label{case11}
L_\alpha\{M\} = L_\beta\{M\} & = & 1 \nonumber\\
L_\gamma\{M\} & = & 0 \qquad (\gamma \not = \alpha,\beta) \,.
\end{eqnarray}
An example is the process $nn\leftrightarrow ppe\mu$,
or simply $WW \to e\mu$.
For this case, 
\begin{equation}
\label{case11amp}
{\cal M}  = \sum_i B_{\alpha\beta i} f_i \,,
\end{equation}
and it is easy to see that
\begin{eqnarray}
\Big| {\cal M} \Big|^2 = \sum_{ij} s_{\alpha ij} s_{\beta ij} f_i f_j^* \,.
\end{eqnarray}
The interference terms are
\begin{equation}
\delta_{\alpha\beta ij} = s_{\alpha ij} s_{\beta ij} \,,
\end{equation}
and the $CP$ violating effects are proportional to 
\begin{equation}
\mbox{Im}\;\delta_{\alpha\beta ij} = (\mbox{Re}\;s_{\alpha ij} )
(\mbox{Im}\;s_{\beta ij}) + (\alpha \leftrightarrow \beta) \,.
\end{equation}
Therefore, in these processes
the $CP$-violating effects due to a non-zero $\mbox{Im}(s_{\beta ij})$
or $\mbox{Im}(s_{\alpha ij})$ vanish unless the real
part of $\mbox{Re}(s_{\alpha ij})$
or $\mbox{Re}(s_{\beta ij})$ is non-zero, respectively.

\subsection{$\Delta L = 0$ processes}
We consider two possibilities, depending on whether two flavors change
(in an opposite way), or three flavors change. 

\subsubsection{Two flavors change}
\label{subsec:main}
In this case the amplitude is such that, for two given flavors
$\alpha,\beta$,
\begin{eqnarray}
\label{case1-1}
L_\alpha\{M\} = - L_\beta\{M\} & = & 1 \,, \nonumber\\*
L_\gamma\{M\} & = & 0 \qquad (\gamma \not = \alpha,\beta) \,.
\end{eqnarray}
The simplest form of an amplitude for this type of process is
\begin{equation}
{\cal M} = \sum_i A_{\alpha\beta i}f_i \,,
\label{M1-1}
\end{equation}
and it corresponds to, for example, processes such as $\mu + p
\to e + p$. In this case,
\begin{equation}
\left|{\cal M} \right|^2 = \sum_{ij} t_{\alpha i\beta j} f_i f^\ast_j \,,
\end{equation}
and the $CP$ violating effects are proportional to
\begin{equation}
\mbox{Im}\;(t_{\alpha i\beta j}) = 
(\mbox{Im}\;s_{\alpha ij}) (\mbox{Re}\;s_{\beta ij} )-
(\alpha \leftrightarrow \beta) \,,
\end{equation} 
where we have used Eq.\ (\ref{tsrelation}). Once again, this
implies that the $CP$-violating effects due to $s_{\alpha i j}$
and $s_{\beta i j}$ disappear unless 
the real part of $s_{\beta i j}$ and $s_{\alpha i j}$ are non-zero, 
respectively.

\begin{figure}
\begin{center}
\begin{picture}(100,70)(0,-15)
\ArrowLine(0,0)(30,0)
\Text(15,5)[b]{$\ell_\beta$}
\Line(30,0)(70,0)
\Text(50,2)[b]{$\nu_i$}
\ArrowLine(70,0)(100,0)
\Text(85,5)[b]{$\ell_\alpha$}
\Photon(30,0)(50,25){-2}4
\Photon(70,0)(50,25)24
\Photon(50,25)(50,40)23
\ArrowLine(0,40)(30,40)
\Text(15,45)[b]{$\ell_\alpha$}
\Line(30,40)(70,40)
\ArrowLine(70,40)(100,40)
\Text(85,45)[b]{$\ell_\alpha$}
\Text(50,-5)[t]{\large (a)}
\end{picture}
\qquad
\begin{picture}(100,70)(0,-15)
\ArrowLine(0,0)(30,0)
\Text(15,5)[b]{$\ell_\beta$}
\Line(30,0)(70,0)
\Text(50,2)[b]{$\nu_i$}
\ArrowLine(70,0)(100,0)
\Text(85,5)[b]{$\ell_\alpha$}
\Photon(30,0)(30,40)24
\Photon(70,0)(70,40)24
\ArrowLine(30,40)(0,40)
\Text(15,45)[b]{$\ell_\alpha$}
\Line(30,40)(70,40)
\Text(50,45)[b]{$\nu_j$}
\ArrowLine(100,40)(70,40)
\Text(85,45)[b]{$\ell_\alpha$}
\Text(50,-5)[t]{\large (b)}
\end{picture}
\qquad
\begin{picture}(100,70)(0,-15)
\ArrowLine(0,0)(30,0)
\Text(15,5)[b]{$\ell_\beta$}
\Line(30,0)(70,0)
\Text(50,2)[b]{$\nu_i$}
\ArrowLine(100,0)(70,0)
\Text(85,5)[b]{$\ell_\alpha$}
\Photon(30,0)(30,40)24
\Photon(70,0)(70,40)24
\ArrowLine(30,40)(0,40)
\Text(15,45)[b]{$\ell_\alpha$}
\Line(30,40)(70,40)
\Text(50,45)[b]{$\nu_j$}
\ArrowLine(70,40)(100,40)
\Text(85,45)[b]{$\ell_\alpha$}
\Text(50,-5)[t]{\large (c)}
\end{picture}
\end{center}
\caption{Different kinds of diagram for the process
$\ell_\beta\ell_\alpha\to \ell_\alpha\ell_\alpha$. The actual number
of diagrams to be calculated are larger because of crossed diagrams
etc. In diagram (a), the vector boson connected to the upper fermion
line can be the photon or the $Z$-boson. All other vector boson lines
are $W$ lines.}\label{f:muto3e}
\end{figure}
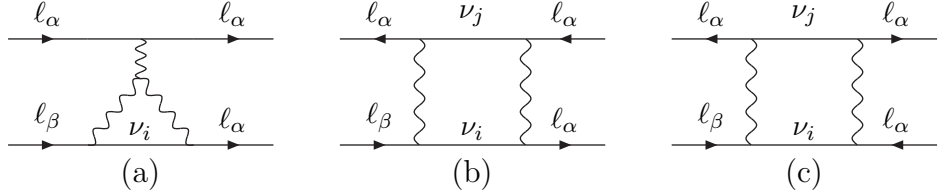
We can consider also terms satisfying Eq.\ (\ref{case1-1}), but built
out of products of the subamplitudes $A,B$.  
Such terms appear, for example,
in the process $\mu + e \to e + e$.  There are two such products
consistent with Eq.\ (\ref{case1-1}), namely $A_{\alpha\beta i}
A_{\alpha\alpha j}$ and $B^*_{\alpha\beta i} B_{\alpha\alpha j}$, and
therefore the total amplitude is of the form
\begin{eqnarray}
{\cal M} 
&=& \sum_i A_{\alpha\beta i} f^a_i 
+ \sum_{ij} \left( A_{\alpha\beta i} A_{\alpha\alpha j} f^b_{ij}
+ B^*_{\alpha\beta i} B_{\alpha\alpha j} f^c_{ij} \right) \nonumber\\*
&=& {\cal M}_a + {\cal M}_b + {\cal M}_c \,.
\label{caseodd}
\end{eqnarray}
Diagrammatically, these three terms correspond respectively to the
diagrams shown in Fig.~\ref{f:muto3e}. It is instructive
to observe the analogy with the process considered previously,
$\mu + p \to e + p$, which involves diagrams like
Fig.~\ref{f:muto3e}a and another that
resembles Fig.~\ref{f:muto3e}b. The latter 
involves the mixing matrix elements from the quark sector,
which are implicitly taken into account in the
structure given in Eq.\ (\ref{M1-1}) by means of the functions $f_i$.

In the case at hand, when we take the absolute square of the amplitude
we obtain, first of all, the terms
\begin{eqnarray}
\Big| {\cal M}_a \Big|^2 &=& \sum_{ij} t_{\alpha i\beta j} f^a_i
f^{a\ast}_j \,, \nonumber\\ 
\Big| {\cal M}_b \Big|^2 &=& \sum_{ijkl} t_{\alpha i\beta k} t_{\alpha j
\alpha l} f^b_{ij} f^{b\ast}_{kl} \,, \nonumber\\ 
{\cal M}_a {\cal M}_b^* &=& \sum_{ijk} t_{\alpha i\beta j} |V_{\alpha
k}|^2 f^a_i f^{b\ast}_{jk} \,,
\end{eqnarray}
which can be expressed in terms of the $t$-type rephasing
invariants alone since the fermion lines conserve total
lepton number. This is not true for the remaining terms
\begin{eqnarray}
{\cal M}_a {\cal M}_c^* &=& \sum_{ijk} s_{\beta ij}^* s_{\alpha ik}
s_{\alpha jk} f^a_{i} f^{c\ast}_{jk} \,,
\nonumber\\ 
{\cal M}_b {\cal M}_c^* &=& \sum_{ijkl} s_{\beta ik}^* s_{\alpha il}
s_{\alpha jk}^* s_{\alpha jl} f^b_{ij} f^{c\ast}_{kl} \,,
\nonumber\\ 
\Big| {\cal M}_c \Big|^2 &=& \sum_{ijkl} s_{\beta ik}^* 
s_{\alpha ik}^* (s_{\alpha jl})^2 f^c_{ij} f^{c\ast}_{kl} \,.
\end{eqnarray}
It is interesting to see that the interference terms that appear in
${\cal M}_a {\cal M}_c^*$,
\begin{eqnarray}
\delta^{(ac)}_{\alpha\beta ijk} \equiv s_{\beta ij}^* s_{\alpha ik}
s_{\alpha jk} \,,
\end{eqnarray}
contain an odd number of $s$-invariants, in contrast with the other terms. 
In particular, their imaginary part 
\begin{eqnarray}
\mbox{Im}\; \delta^{(ac)}_{\alpha\beta ijk} = - \mbox{Im}\; (s_{\beta
ij}) \mbox{Im}\; (s_{\alpha ik}) \mbox{Im}\; (s_{\alpha jk}) +
\mbox{other terms} \,,
\label{imimim}
\end{eqnarray}
do not involve the real part of any $s$-invariants directly.
On the other hand, if the imaginary part of the
$s$-invariants vanish, then the imaginary part of all the interference terms
certainly go to zero, as it should be. The distinctive feature 
of the amplitude given in Eq.\ (\ref{caseodd}) 
is that, as we have seen, it leads to
interference terms that contain an odd number of $s$-invariants.

We can consider other processes where two flavors change, keeping the
total lepton number conserved. For example, consider the case 
\begin{eqnarray}
\label{case2-2}
L_\alpha\{M\} = - L_\beta\{M\} & = & 2 \,, \nonumber\\*
L_\gamma\{M\} & = & 0 \qquad (\gamma \not = \alpha,\beta) \,,
\end{eqnarray}
which corresponds to processes of the type $\ell_\beta\ell_\beta \to
\ell_\alpha\ell_\alpha$ or, by crossing, to
processes such as muonium-antimuonium oscillations.
The diagrams for this case are similar to those given in
Fig.~\ref{f:muto3e}b,c, where the incoming $\ell_\alpha$ line has to be
replaced by an incoming $\ell_\beta$ line. The corresponding 
amplitude is
\begin{eqnarray}
{\cal M} (\ell_\beta\ell_\beta\to \ell_\alpha\ell_\alpha) 
&=& \sum_{ij} \left( A_{\alpha\beta i} A_{\alpha\beta j} f^b_{ij}
+ B^*_{\beta\beta i} B_{\alpha\alpha j} f^c_{ij} \right) \nonumber\\
&=& {\cal M}_1 + {\cal M}_2 \,,
\end{eqnarray}
and squaring we obtain
\begin{eqnarray}
\Big| {\cal M}_1 \Big|^2 &=& \sum_{ijkl} t_{\alpha i \beta k}
t_{\alpha j \beta l} f^b_{ij} f^{b\ast}_{kl} \,, \nonumber\\ 
{\cal M}_1 {\cal M}_2^* &=& \sum_{ijkl} s_{\alpha il} s_{\alpha jl}
s_{\beta ik}^* s_{\beta jk}^* f^b_{ij} f^{c\ast}_{kl} \,, \nonumber\\
\Big| {\cal M}_2 \Big|^2 &=& \sum_{ijkl} (s_{\alpha jl})^2 (s_{\beta
ik}^*)^2 f^c_{ij} f^{c\ast}_{kl} \,. 
\end{eqnarray}
It is easy to see that the $CP$ violating terms contain 
products of the real and imaginary parts of the $s$-invariants,
and therefore the possible effects of a particular element $s_{\alpha ij}$
vanish unless the real part of a corresponding element is non-zero.

\subsection{Three flavors change}\label{s:-1-12}
This case corresponds to an amplitude such that for three given
flavors $\alpha,\beta,\gamma$, we have
\begin{eqnarray}
\label{case-1-12}
L_\alpha\{M\} = L_\beta\{M\} & = & - 1 \nonumber\\
L_\gamma\{M\} & = & 2 \nonumber\\
L_\delta\{M\} & = & 0 \qquad (\delta \not = \alpha,\beta,\gamma) \,.
\end{eqnarray}
Examples of such processes are $\mu + e \to \tau + \tau$ and
$\mu + \tau \to e + e$.
The products that are consistent with Eq.\ (\ref{case-1-12}) are
$A_{\gamma\alpha i}A_{\gamma\beta j}$ and $B^*_{\alpha\beta i}
B_{\gamma\gamma j}$, and therefore the amplitude is
of the form
\begin{eqnarray}
{\cal M} = \sum_{ij} \left( A_{\gamma\alpha i}A_{\gamma\beta j} f^a_{ij}
+ B^*_{\alpha\beta i} B_{\gamma\gamma j} f^b_{ij} \right) \,.
\end{eqnarray}
In this case also, the $CP$ violating terms appear as products of real
parts and imaginary parts of the $s$-invariants.

\subsection{Higher order corrections}
\label{subsec:radcorr}
So far we have found that we can divide the processes into two
classes.  In the first class, the $CP$-violating terms in the squared
amplitude consist of products of the real and the imaginary parts of
the $s$-invariants. For processes in this class, the $CP$-violating
effects of a given invariant element $s_{\alpha ij}$ vanishes unless
the real of some corresponding element is non-zero.  In the second
class, which contains, for example, the process
$\ell_\beta\ell_\alpha\to \ell_\alpha\ell_\alpha$ considered in
Section\ \ref{subsec:main}, the vanishing of all the $CP$-violating
terms requires that the imaginary parts of the invariants be zero. As
already hinted at in the argument that lead to Eq.\ (\ref{imimim}),
the feature that distinguishes these two classes is the following.  In
the first class the amplitude is such that the interference terms
consist of products of only an even number of $s$-invariants, while in
the second class some interference terms contain an odd-number of
them.

In fact, any process for which the corresponding amplitude is
of such a form that, when it is squared, the interference terms can be
expressed as products of an even number of $s$-invariants
would fall in the first class. 
However, even for this class of process, in general, 
this property is lost when higher order diagrams are taken into account.
We now show that the higher order terms in the amplitude 
lead to interference terms that contain an odd number
$s$-invariants, and hence $CP$ is strictly conserved in those
processes only if the imaginary part of the relevant invariant
elements vanish.

\begin{figure}
\begin{center}
\begin{picture}(150,60)(0,-15)
\ArrowLine(30,0)(0,0)
\Text(15,5)[b]{$\ell_\alpha$}
\Line(30,0)(60,0)
\Text(45,5)[b]{$\nu_j$}
\ArrowLine(90,0)(60,0)
\Text(75,5)[b]{$\ell_\gamma$}
\Line(90,0)(120,0)
\Text(110,5)[b]{$\nu_k$}
\ArrowLine(120,0)(150,0)
\Text(135,5)[b]{$\ell_\beta$}
\Photon(30,0)(30,40)24
\Photon(90,0)(90,40)24
\PhotonArc(90,0)(30,180,360){2}{8.5}
\end{picture}
\end{center}
\caption{A higher order correction to diagram (b) of Fig.\ \ref{f:blocks}.
\label{fig:radcorrB}
}
\end{figure}
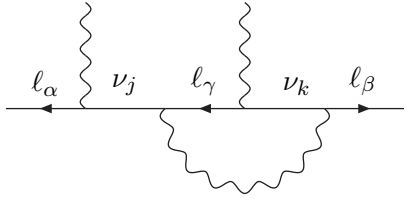

As an example, we reconsider the processes such as $WW\to e\mu$ which were
discussed in Section\ \ref{subsec:case2}.  In addition to the term
$B_{\alpha\beta i}$ considered in Eq.\ (\ref{case11amp}), the amplitude
can contain the term
\begin{equation}
\label{radcorrtoB}
\sum_{\gamma jk}A_{\alpha\gamma j}B_{\gamma\beta k}f_{\gamma jk} 
\end{equation}
which is consistent with the condition given in Eq.\ (\ref{case11}).
Such a term can arise, for example, from the diagram given in Fig.\
\ref{fig:radcorrB}.
In general, whenever the subamplitude $A$ or $B$ appears in an
expression, we can make the replacements 
\begin{eqnarray}
A_{\alpha\beta i} &\to& B_{\alpha\gamma j} B^\ast_{\gamma\beta k}\,,\nonumber\\
B_{\alpha\beta i} &\to& A_{\alpha\gamma j} B_{\gamma\beta k}
\end{eqnarray}
to obtain a higher order term in the amplitude,
because both sides carry the same $L_\delta$ for all $\delta$. 

When the total amplitude, containing the terms given in
Eqs.\ (\ref{case11amp}) and (\ref{radcorrtoB}), is squared,
there will be an interference term
\begin{equation}
\sum_{\gamma ijk}\delta_{\alpha\beta\gamma ijk}f^\ast_i f_{\gamma jk}
\end{equation}
where
\begin{eqnarray}
\delta_{\alpha\beta\gamma ijk} & = & B^\ast_{\beta\alpha i}
A_{\alpha\gamma j}B_{\gamma\beta k} \nonumber\\
& = & s_{\beta ki}s_{\gamma kj}s_{\alpha ji} \,. 
\end{eqnarray}
Therefore, the $CP$ violating part of the amplitude involves products
of the type $\mbox{Im}(s_{\beta ki}) \mbox{Im}(s_{\gamma kj})
\mbox{Im}(s_{\alpha ji})$, which do not necessarily vanish even if
some of the $\mbox{Re}(s_{\alpha ik})$ are zero.

Similar examples can be constructed for other processes discussed
here. For example, consider the process discussed in
Sec.~\ref{s:-1-12}. The higher order corrections would
include the following terms in the amplitude
\begin{eqnarray}
\sum_{\delta ijk} A_{\delta\alpha k} B^\ast_{\beta\delta i} 
B_{\gamma\gamma j} f_{\delta ijk} \,.
\end{eqnarray}
Here the interference terms will also contain products of the imaginary
parts several $s$-invariants, similarly to the previous case.

\section{Conclusions}
\label{sec:conclusions}
We have presented a variety of examples to illustrate how the mixing
and $CP$ violating parameters of the lepton sector appear in rephasing
invariant combinations in the squared amplitudes for various types of
processes.  As we have already discussed in Section\
\ref{subsec:radcorr}, in general the $CP$-violating terms contain
products of an odd number of rephasing invariants, and hence $CP$ is
strictly conserved only if the imaginary part of the relevant
invariant elements vanish.  However, there are processes in which the
leading order $CP$-violating terms contain products of an even number
of the invariants, and hence the $CP$-violating effects may vanish if
the real part of the appropriate rephasing invariants are zero. 
Nevetheless, as we have seen, this characteristic is lost when higher order
corrections are included.

It is well known that, in the presence of Majorana neutrinos, there
are additional $CP$ violating parameters as compared to, for example,
the quark sector.  While it is sometimes believed that these extra
parameters appear only in lepton number violating processes
\cite{Schechter:1981gk}, we have shown that they can, and in fact they
do, appear in lepton number conserving processes as well. The true
condition for the occurrence of these parameters in a given process
seems to be the violation of lepton number on any fermion line in the
corresponding diagrams, and not necessarily that total lepton number
be violated by the process as a whole.  In processes that conserve the
total lepton number, there are in general diagrams in which the
individual fermion lines change the lepton number, but do so in such a
way that the changes between different lines cancel in the overall
diagram.  The interference terms produced by such diagrams would
contain the extra $CP$-violating parameters that exist due to the
Majorana nature of the neutrinos.

\paragraph*{Acknowledgment}
The authors thank S. T. Petcov and L. Wolfenstein for
discussions. This work has been  partially 
supported (JFN) by the U.S. National Science Foundation Grant PHY-9900766. 


\end{document}